\documentclass[USenglish,
10pt,
twocolumn
]{article}
\newcommand{\shortcite}[1]{\cite{#1}}
\usepackage[latin1]{inputenc}
\usepackage{palatino}
\usepackage{vmargin}
\usepackage
{graphicx}
\usepackage{epstopdf}

\usepackage{setspace}
\usepackage{amsmath}
\usepackage{amsthm}
\usepackage{amssymb}
\usepackage{enumerate}

\setpapersize{A4}
\setmarginsrb{2.5cm}{0.5cm}{2.5cm}{3.5cm}{1.5cm}{1cm}{1.5cm}{1.5cm}

\newcommand{\e}[1]{\emph{#1}}
\newcommand{\Cs}{\mathbf{C}} 

\newcommand{\Ss}{\mathbf{S}}

\newcommand{\ws}{{\wedge\star}}
\newcommand{\REL}{\mathcal{R}}

\newcommand{\T}[1]{\texttt{#1}}

\newtheorem*{definition*}{Definition}

\title{Lattices for Dynamic, Hierarchic \& Overlapping Categorization: the Case of Epistemic Communities
\author{Camille Roth\footnote{CREA (Center for Research in Applied
    Epistemology), CNRS/Ecole Polytechnique, 1 rue Descartes, 75005 Paris, France.
    Corresponding author: \emph{roth@shs.polytechnique.fr}} ,
   Paul Bourgine\footnotemark[\value{footnote}]
  }}
\begin{document}
\maketitle 
\begin{abstract}
We present a method for hierarchic categorization and  taxonomy evo\-lu\-tion description.  We focus on the structure of epistemic communities (ECs), or groups of agents sharing common knowledge concerns. Introducing a formal framework based on Galois lattices, we  categorize ECs in an automated and hierarchically structured way and propose criteria for selecting the most relevant epistemic communities --- for instance, ECs gathering a certain proportion of agents and thus prototypical of major fields. This process produces a manageable, insightful taxonomy of the community. Then, the longitudinal study of these static pictures makes possible an historical description. In particular, we capture stylized facts such as field progress, decline, specialization, interaction (merging or splitting), and paradigm emergence. The detection of such patterns in social networks could fruitfully be applied to other contexts.
\newline

\end{abstract}

%\setlength{\columnseprule}{0.4pt}
%\begin{multicols}{2}
  
\section*{Introduction}

A taxonomy is a hierarchical structuration of things into categories. It is a fundamental concept for understanding the organization of groups of items sharing some properties. Taxonomies are useful in many different disciplinary fields: in biology for instance,  where classification of living beings has been a recurring task \cite{%simp:real,
whit:newc}; in cognitive psychology for modelling categorical reasoning \cite{rosc:cogn}; as well as in ethnography and anthropology with e.g. folk taxonomies \cite{berl:ethn,lope:tree}. 
In this paper, we focus on the structure of knowledge communities. More precisely, we aim at rebuilding an evolving  taxonomy of groups of agents sharing common knowledge concerns, or \emph{epistemic communities} \cite{cowa:expl,haas:intr}. 

While taxonomies have initially been built using a subjective approach, the focus has slipped to formal and statistical methods \cite{soka:nume}. 
Simultaneously, along with the massive development of  informational content, dealing with and ordering categories in an automated fashion has become a central issue in data mining and related fields %\cite{chen:fuzz, koba:info}. %[REF: {\em Srikant R. Agrawal R. Mining Generalized Association Rules 1995}]
\cite{jain:data}.
Many different techniques indeed have been proposed for producing and representing categorical structures  %\cite{jain:data}
 including, to cite a few, %in particular
 hierarchical clustering \cite{john:hier}, graph theory-based techniques \cite{newm:dete}, formal concept analysis \cite{will:rest}, information theory \cite{leyd:stat}, {\em Q}-analysis \cite{atki:math}, blockmodeling \cite{bata:gen2}, neural networks \cite{koho:soms}, association mining \cite{srik:mini}, and dynamic exploration of taxonomies \cite{sacc:dyna%,woll:meth
}. 

In scientometrics in particular, categorization has been applied to  scientific community representation, using \emph{inter alia} multidimensional scaling in association with co-citation data \cite{kreu:coci,mcca:coci} or other co-occurrence data \cite{noyo:moni}, in order to produce two-dimensional cluster mappings.

Among this profusion of clustering methods, %when such a clustering method is being used to the end of building a taxonomy, it is mostly yielding simple a tree-like structure called ``dendrogram''. 
taxonomy building itself has yet been poorly investigated; arguably, taxonomy evolution during time has been fairly neglected.
Our intent here is to address both topics. At the same time, we intent to deal with items belonging to multiple categories or having diverse paradigmatic statuses. 
We therefore propose a method based on Galois lattices \cite{birk:latt,roth:epis} to represent a relevantly reduced view of such a taxonomy.
Then, we describe the community taxonomy in an historical perspective by studying the evolution of these static pictures. 
In particular, we rebuild stylized facts relating to epistemic evolution. These facts consist of field progress or decline, %through the variation in the number of agents of the corresponding epistemic community, (ii)
 field scope  enrichment or impoverishment,  %namely the merging of several existing fields into a more specific subfield, or the regrouping of several fields into a more general one, or the splitting of a field into various subfields, 
 and  field interaction (merging or splitting). This would be useful for disciplines such as history of science and scientometrics. It would also provide agents with automated methods to know the structure of the community they are evolving in.

In section \ref{sec:formal} we introduce  the formal framework needed for representing epistemic community taxonomy using Galois lattices. Section \ref{sec:category} describes how to build  recuced taxonomies, and section \ref{sec:taxonomy} adresses their evolution. A case study is investigated in section \ref{sec:case}, followed by a general discussion in section \ref{sec:discussion}.

\section{Formal framework}
\label{sec:formal}
%In this section we define.\footnote{Interested readers may find more details about this section in \cite{roth:epis}.}

\subsection{Epistemic communities}

\paragraph{Binary relation, intent, extent} We introduce the notion of \emph{epistemic community}. %to describe the taxonomy of knowledge communities, 
In the litterature  \cite{cowa:expl,dupo:orga,haas:intr}, an epistemic community is a group of agents sharing a common set of subjects, topics, concerns, sharing a common goal of knowledge creation. 
In order to use this notion, we first have to bind agents to semantic items, or \emph{concepts}. 

 To this end, we consider a binary relation $\REL$ between an agent set $\Ss$ and a concept set $\Cs$. $\REL$ expresses any kind of relationship between an agent $s$ and a concept $c$. The nature of the relationship depends on the hypotheses and the empirical data. In our case, the relationship represents the fact that $s$ used $c$ in some article. 

We may thus introduce two fundamental notions: the \emph{intent} and the \emph{extent}. The \emph{intent} $S^{\wedge}$ of an agent set $S$ is the set of concepts that is being used by every agent in $S$. Similarly, the \emph{extent} $C^{\star}$ of a concept set $C$ is the set of agents using every concept in $C$. 
%Therefore, an EC based on an agent set $S$ is the largest agent set with the same intent as $S$. 

\paragraph{Epistemic community} 
We then adopt  the following definition: \emph{an epistemic community (EC) is the largest set of agents sharing a given concept set}. 
Accordingly, an \emph{EC based on an agent set} is the EC of its concept set. Such EC is the largest agent set having in common  the same concepts  as the initial agent set. In other words, taking the EC of a given agent set extends it to the largest community sharing its concepts.
Notice that this notion strongly relates to structural equivalence \cite{lorr:stru}, with ECs being groups of agents linking equivalently to some concepts.

The EC based on an agent set $S$ is therefore the largest agent set with the same intent as $S$. It is then obvious that this largest set is the extent of the intent of $S$, or $S^{\ws}$. Thus, the operator ``$\ws$'' yields the EC of any agent set. Notice that we can similarly define an \emph{EC based on a concept set} as the largest set of concepts sharing a given agent set. Here, one starts with a concept set and seeks to know its corresponding EC using the operator ``$\star\wedge''$. The EC based on a concept set $C$ is the same as that based on an agent set $S=C^\wedge$. Hence, in the remainder of the paper we will equivalently denote an EC by its agent set $S$, its concept set $C$ or the couple $(S,C)$.

\subsection{Building taxonomies}

\paragraph{Community structure and lattices}
Assuming that knowledge communities are structured into fields and subfields, the raw set of all ECs  is not sufficient to build a taxonomy: we need to hierarchize it. The canonical approach for representing  and ordering categories consists of trees, which render Aristotelian taxonomies. In a tree, categories are nodes, and sub-categories are child nodes of their unique parent category. A major drawback of such a taxonomy lies in its ability to deal with objects belonging to multiple categories. In this respect, the \emph{platypus} is a famous example: it is a mammal and a bird at the same time.  Within a tree, it has to be placed either under the branch \emph{``mammal''}, or the branch \emph{``bird''}. Another problem is that trees make the representation of paradigmatic categories  extremely unpractical. Paradigmatic classes are categories based on exclusive (or orthogonal) rather than hierarchical features \cite{voge:taxi}: for instance \emph{urban} vs. \emph{rural}, \emph{Italy} vs. \emph{Germany}.  In a tree, \emph{``rural Italy''} has to be a subcategory of either \emph{rural} or \emph{Italy}, whereas there may well be no reason to assume an order on the hierarchy and a redundancy in the differenciation. 

\begin{figure}\begin{center}\includegraphics[width=6.23cm%9.83cm%0.82\linewidth%, trim=-10 0 5 0
]{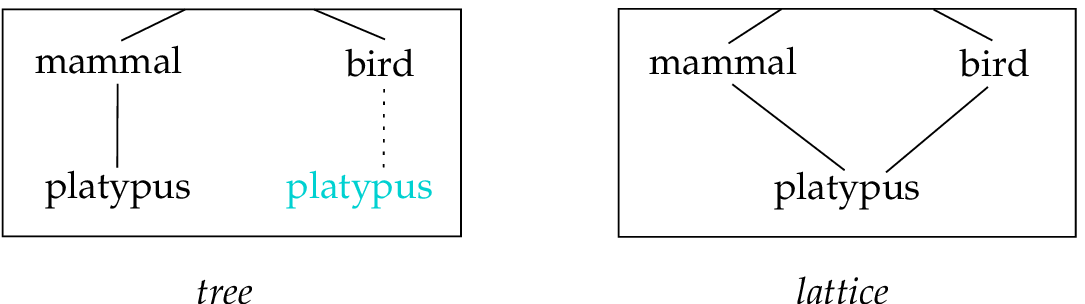}
\vspace{0.45cm}\hrule\vspace{0.45cm}
\includegraphics[width=7.45cm%0.98\linewidth%6.9cm%, trim=10 0 5 0
]{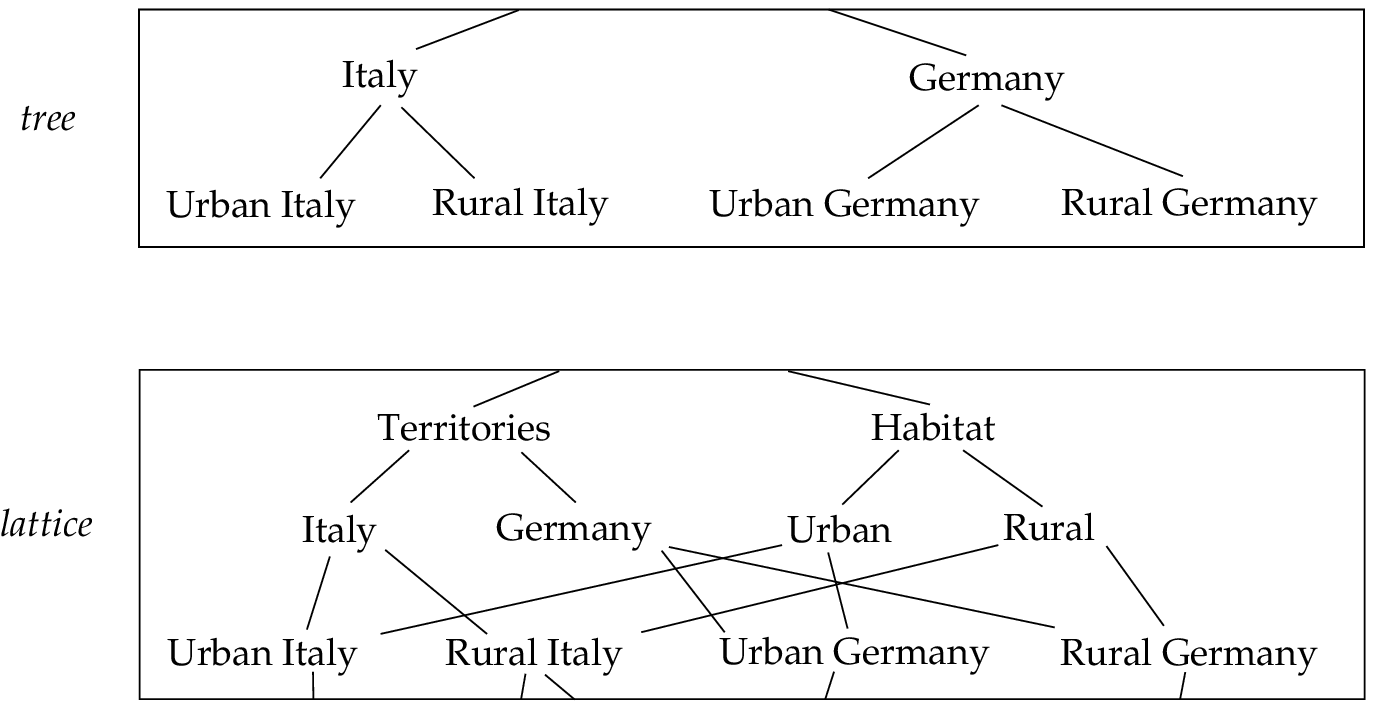}\end{center}\caption{\label{fig:paradigmatic} Trees \hbox{vs.} lattices. \emph{Top}, multiple categories: in a tree, the \emph{platypus} needs  either to be affiliated with \emph{mammal} or \emph{bird}, or to be duplicated in each category --- in a lattice, this multiple ascendancy is effortless. \emph{Bottom}, paradigmatic taxonomies: in a tree, a paradigmatic distinction (e.g. territories vs. habitat types) must lead to two different levels --- in a lattice, the two paradigmatic notions may well be on the same level, leading to mixed sub-categories.}
\end{figure}

A straightforward way to improve the classical tree-based structure is a lattice-based structure, which allows \emph{category overlap}  representation. Technically, a lattice is a partially-ordered set such that given any two elements $l_1$ and $l_2$, the set $\{l_1,l_2\}$ has  a least upper bound (denoted by $l_1\vee l_2$ and called ``join'') and a greatest lower bound (denoted by $l_1\wedge l_2$ and called ``meet'').
In a lattice, the \emph{platypus} may simply be the sole member of the joint category \emph{``mammal-bird''}, with the two parent categories \emph{``mammal''} and \emph{``bird''}.  The \e{``mammal-bird''} category is \e{``mammal''}$\wedge$\e{``bird''}, \hbox{i.e.} \e{``mammal''}-meet-\e{``bird''}. The parent category (``animal'') is \e{``mammal''}$\vee$\e{``bird''}, or \e{``mammal''}-join-\e{``bird''}.
Besides, lattices may also contain different kinds of paradigmatic categories at the same level -- see Fig. \ref{fig:paradigmatic}.

\begin{figure}\begin{center}
\includegraphics[width=7.22cm%0.95\linewidth
, trim=-10 0 5 0]
{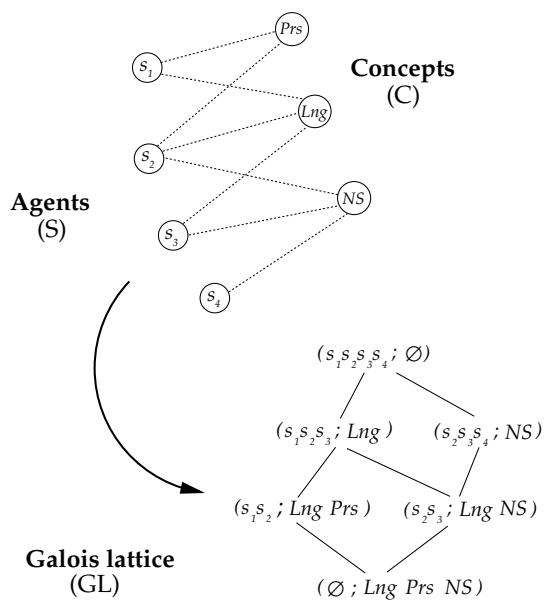}
\caption{\label{fig:galoisexemple} %\small
Example of binary relation with 4 agents and 3 concepts, \emph{prosody} (Prs), \emph{linguistics} (Lng) and \emph{neuroscience} (NS) --- below, the corresponding Galois lattice (6 ECs); lines indicate  hierarchic relationships: from top (most general) to bottom (most specific); ECs are represented as a pair (extent, intent) = $(S, C)$ with $S^\wedge=C$ and $C^\star=S$.}
\end{center}
\end{figure}

\paragraph{Galois lattices}
We hence argue that a lattice replaces efficiently and conveniently trees for describing taxonomies, and particularly knowledge community structure.\footnote{We will not consider graded categories like fuzzy categories \cite{zade:fuzz} and thick categories (such as locologies \cite{bart:logi}).} Therefore, we define the following \emph{partial order} between ECs:  an EC is a subfield of a field \emph{if} its intent is more precise than that of the field; in other words, if the concept set of the subfield contains that of the field.  %This defines a \emph{partial order} between ECs.
Provided with this order, the Galois lattice is the ordered set of all ECs \cite{birk:latt}, as shown on Fig. \ref{fig:galoisexemple}. An EC closer from the top is more general: the hierarchy reproduces a generalization/specialization relationship. Besides, joint categories are descendants of several ECs (they form ``diamond bottoms'').

In this respect, GLs are a very natural tool for building taxonomic lattices from a binary relation between agents and concepts. More generally, it is worth noting that we can replace \emph{authors} with \emph{objects}, and \emph{concepts} with \emph{properties}. This yields a generic method for producing a comprehensive taxonomy of any field where categories can be described as a set of items sharing equivalently some property set. This has been indeed a useful application of GLs in artificial intelligence (as ``Formal Concept Analysis'') \cite{gant:algo,godi:met2,will:rest}, and has been investigated as well  in mathematical sociology  recently \cite{bata:gen2,free:usin}. 
However, a serious caveat of GLs is that they may grow extremely large and therefore become very unwieldy. Indeed, even for a small number of agents and concepts, GLs contain often significantly more than several thousands of ECs.
In the next sections, we show how to use GLs both to produce a \emph{manageable} taxonomy and to monitor its \emph{evolution}.

\section{Community selection}\label{sec:category}

\subsection{Rationale}\label{sec:falsepositives}
\begin{figure}\begin{center}
\includegraphics[width=6.46cm%0.85\linewidth%, trim=65 50 0 5
]{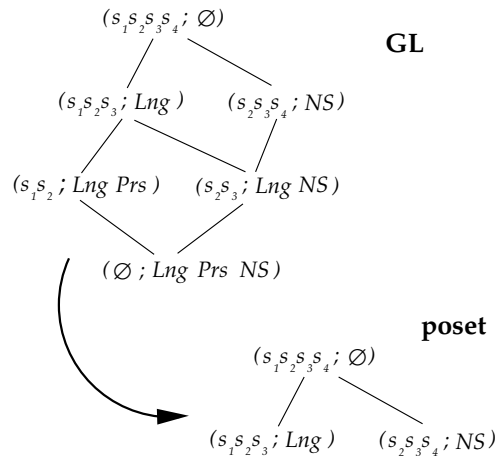}
\end{center}
\caption{\label{fig:poset} From the original GL to a selected poset.}
\end{figure}

%Our aim is to rebuild the taxonomy of a given community 
GLs are thus usually very large and in a dynamic perspective, it is significantly harder to track a series of GLs than just examining a static lattice. 
Therefore, considering only useful and meaningful patterns instead of manipulating whole lattices becomes utmost crucial. This means selecting from a possibly huge GL which ECs are relevant to taxonomy rebuilding, and excluding a large number of irrelevant ECs that could blur the picture of the community.  
In other words, we consider a partial, manageable view of the whole GL which we choose in order to  reflect the most significant part and patterns of the taxonomy. Formally, the partial view is not anymore a lattice as defined previously: it is a partially-ordered set, or \emph{poset}; nonetheless it overlays on the lattice structure and still enjoys the taxonomical properties we are interested in (see Fig. \ref{fig:poset}). For the sake of clarity, we will name  \emph{``partial lattice''} such a poset.%, keeping in mind that is not  a lattice formally.

\paragraph{Selection preferences} This selection process has so far been an underestimated topic in the study of GLs, with an important part of the effort focused on GL computation and representation \cite{dick:ares,ferr:file,godi:met2,kuzn:comp}. Nevertheless, some authors insist on the need for semantic interpretations and approximation theories in order to cope with GL combinatorial complexity \cite{duqu:stru,vand:comp}.
In our case, we need to specify selection preferences, i.e. which kind of ECs are relevant for a concise taxonomy description. This implies for instance to keep those that correspond to basic-level categories, in Rosch's sense \cite{rosc:cogn}.  At first, we would certainly focus on the largest ECs: if a set of properties, attributes or concepts corresponds to a field, one can expect that the corresponding extent is of a significant size. Thus one would focus on high-size closed sets, while ignoring either too small or too specific closed sets.

\begin{figure}\begin{center}
\includegraphics[width=7.14cm%1.15\linewidth
%, trim=65 50 0 5
]{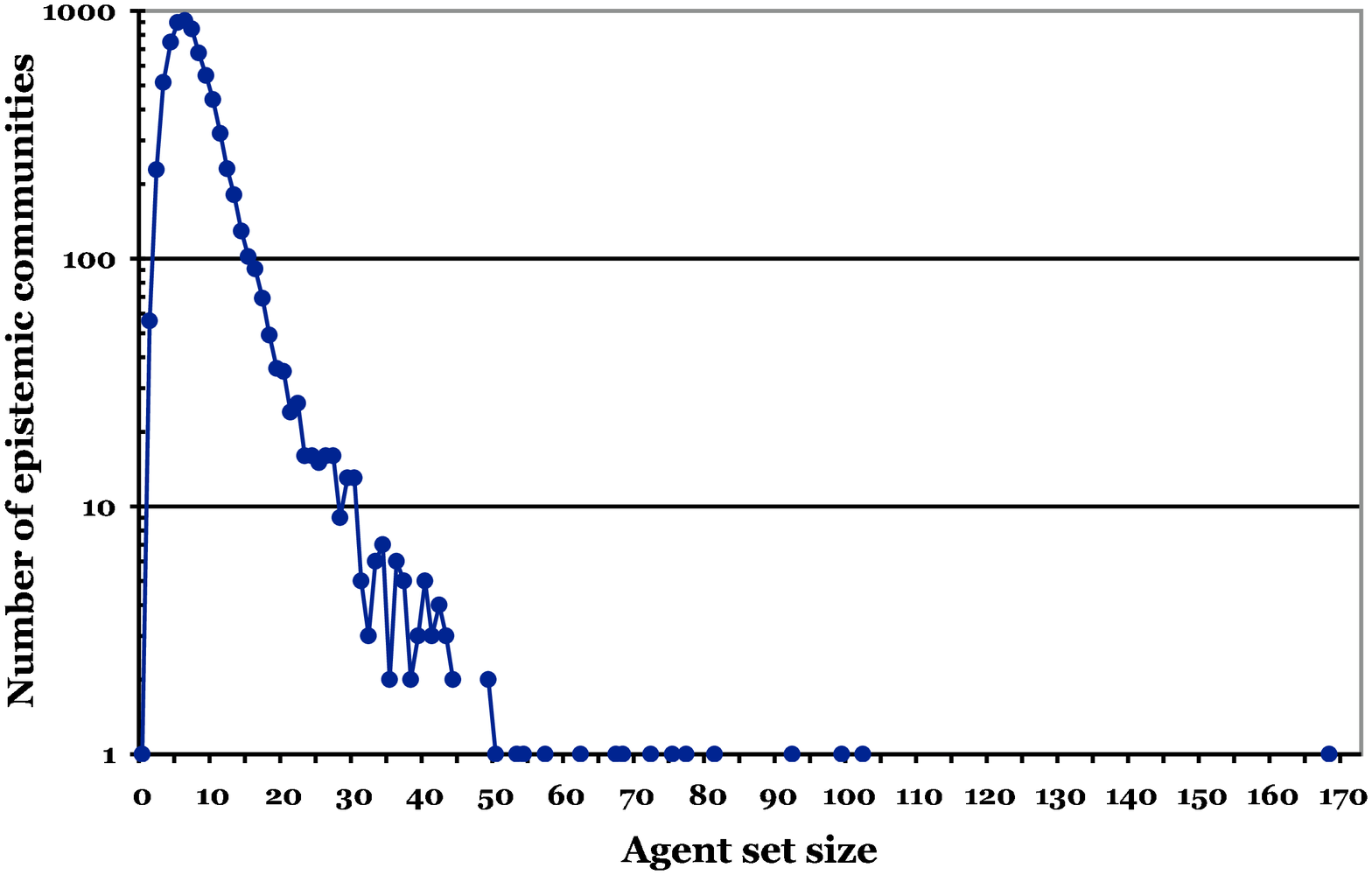}
\caption{\label{fig:distrib} %\small
Raw distribution of epistemic community sizes, in a typical GL calculated for a relationship between 250 agents and 70 concepts.}
\end{center}
\end{figure}
We previously used the size criterion in a first approach on epistemic community categorization through GLs  \cite{roth:epis}. Since fields tend to be made of large groups of agents, and also because  a GL mostly consists of small communities (see Fig.\ref{fig:distrib}), size proved to be a segregating and efficient criterion, categorizing a large portion of the whole community --- however still an unsufficient criterion. Indeed, using only this criterion may be over-selective or under-selective, notably in the following cases:
\begin{itemize}
\item \emph{Small yet significant sets}. One should not pay attention to very small closed sets, for instance those of size one or two: in general they cannot be considered representative of  any particular EC. There is thus a pertinent threshold for the size criterion. However, this may still exclude some small ECs that could actually be relevant, notably those prototypical of a minority community. If so, some other criteria might apply as well: 

(i) such ECs indeed, while being small, are unlikely to be subsets of other ECs and are more likely to be located in the surroundings of the lattice top;

(ii) alternatively, they may be unusually specific with respect to their position in the lattice; % (e.g., referring to 3 concepts while only at distance 2 from the top); 

(iii) finally, being outside the mainstream may make them less likely to mix with other ECs, thus having fewer descendants. 

\item {\em Large yet less significant sets}. 
Large contingent ECs may augment the GL uselessly. This is the case: 

(i) when two ECs are large: it is likely that their intersection exists and has fortuitously a significant size --- we could  discriminate ECs whose size is not significant enough with respect to their smallest ascendant. 

(ii) when empirical data fails to mention that some agents are linked to some properties: two or more very similar ECs appear where only one exists in the real world\footnote{Indeed, let $s_1$, $s_2$, $s_3$, $s_4$ and $s_5$ work on $c_1$, $c_2$, $c_3$, $c_4$ and $c_5$, in reality. Suppose now that some data for $s_5$ is missing and that we are ignorant of the fact that $s_5$ works on $c_5$. Then there will be two distinct communities: $(\{s_1,\,s_2,\,s_3,\,s_4\},\:\{c_1,\,c_2,\,c_3,\,c_4,\,c_5\})$ and $(\{s_1,\,s_2,\,s_3,\,s_4,\,s_5\},\:\{c_1,\,c_2,\,c_3,\,c_4\})$, which cover a single real EC.}  --- we could avoid this duplicity by excluding  ECs whose size is too close to that of their smallest ascendant.

\end{itemize}

\subsection{Selection methodology}

\paragraph{Extending preferences and criteria}

Hence, a\-gent set si\-ze does not matter alone and selection preferences cannot be based on size only.
For instance, small ECs distant from the top are likely to be irrelevant, and certainly the most uninteresting ECs are the both smaller and less generic ones. 
To keep small meaningful ECs and to exclude large unsignificant ones, some more  criteria are required to design the above preferences.
For a given epistemic community $(S,\,C)$, we may propose the following criteria:% taking into account the previous considerations: %along with their respective variables:
\begin{enumerate}
\item size (agent set size), $|S|$;
\item  level (shortest distance to the top\footnote{We take here the shortest length of all paths leading to the top EC $(\Ss,\,\emptyset)$ (the whole community). Indeed, paths from a node to the top are not unique in a lattice; we could also have chosen, for instance, the average lengths of all paths.}), $d$;
\item  specificity (concept set size), $|C|$;
\item  sub-communities (number of descendants), $n_d$;
\item contingency\,/\,relative size (ratio between the agent set and its smallest ascendant), $\lambda$.
\end{enumerate}

%NOTER POUR LA THESE: ON PEUT SYMETRISER LES CRITERES.

\paragraph{Selection heuristics}
Then, we design several simple selection heuristics adequately rendering selection preferences. Selection heuristics are functions  attributing a score to each EC by combining these criteria, so that we only keep the top scoring ECs.
We may not necessarily be able to express all preferences through a unique heuristic. Therefore, the selection process involves several heuristics: for instance one function could select large communities, while another is best suited for minority communities. We ultimately keep the best nodes selected by each heuristic (e.g. the 20 top scoring ones). 

Notice that agent set size $|S|$ remains a major criterion and should take part in every heuristic. Indeed, a heuristic that does not take size into account could assign the same score, for example, to a very small EC with few descendants (like those at the lattice bottom) and to a larger EC with as many few descendants (possibly a worthy heterodox community). In other words, given an identical size, heuristics will favor ECs closer to the top, having less descendants, etc. In general we need heuristics that keep the significant upper part of the lattice. Hence distance to the top $d$ is important as well and should be used in many heuristics. 

While we can possibly think of many more criteria and heuristics, we must yet make a selection among the possible selection heuristics, and pick out some of the most convenient and relevant ones.  In this respect, the following heuristics are a \emph{possible} choice:% for involving each criterion and rendering the preferences expressed in the previous paragraphs:
\begin{enumerate}
\item $|S|$ : select large ECs,
\item $\displaystyle\frac{|S|}{d}$ :  select large ECs close to the top,
\item $\displaystyle |S|\frac{|C|}{d}$ : select large ECs unusually specific,
\item $\displaystyle \frac{|S|}{d}%\cdot 
n_d$ : select large ECs close to the top and having few descendants,
\item $\displaystyle \frac{|S|}{d}(\lambda-\lambda^{+})(\lambda^{-}-\lambda)$: select large {non-con\-tin\-gent} ECs close to the top.\footnote{That is, of a moderate size relatively to their parents: \mbox{$\lambda\in [\lambda^-;\lambda^+]$} --- we thus expect to exclude fortuitous EC intersections when $\lambda<\lambda^-$, and duplicate ECs when $\lambda>\lambda^+$.}
\end{enumerate}

Fine tuning these heuristics eventually requires an active feedback from empirical data. For instance, one could prefer to consider only the first heuristics, and accordingly to focus on taxonomies including only large, populated, dominant ECs. Exploring further the adequacy and optimality of the choice and design of these heuristics would also be an interesting task  ---  heuristics yielding \hbox{e.g.} a maximum number of agents for a minimal number of ECs --- however unfortunately far beyond the scope of this paper. We will thus authoritatively keep and combine these few heuristics to build the partial lattice from the original GL, as shown on Fig. \ref{fig:poset}. In any case, correct empirical results with respect to the rebuilding task will acknowledge the validity of this choice.

\section{Taxonomy evolution}\label{sec:taxonomy}
To monitor  taxonomy evolution we  monitor partial lattice evolution. To this end, we create a series of partial lattices from  GLs corresponding to each period, and  we capture some \emph{patterns} reflecting epistemic evolution by comparing successive static pictures. In other words, we proceed to  a longitudinal study of this series. 

Interesting patterns include in particular:
\begin{itemize}
\item {\em progress or decline of a field}: a burst or a lack of interest in a given field;
\item {\em enrichment or impoverishment of a field}: the reduction or the extension of the set of concepts  related to a field;
\item {\em reunion or scission of fields}: the merging of several existing fields into a more specific subfield %or a more general superfield, 
or the scission of various fields previously mixed.
\end{itemize}

In terms of changes between successive partial lattices, the first pattern simply translates into a variation in the population of a given EC: the agent set size increases or decreases.% (Fig. \ref{fig:dynpatterns}--top).

The second pattern reduces in fact to the same phenomenon. Indeed,
suppose \e{``linguistics''} is enriched by \e{``prosody''}, i.e. $\{Lng\}$ is enriched by $\{Prs\}$, thus becoming $\{Lng, Prs\}$. This  means that the population of $\{Lng, Prs\}$ is increasing. Since this EC is still a subfield of $\{Lng\}$, the enrichment of $\{Lng\}$ by $\{Prs\}$  translates into an increase of  its subfield. Similarly, the decrease of $\{Lng, Prs\}$ would indicate an impoverishment of the superfield $\{Lng\}$.\footnote{More formally, say a field $(S, C_1)$ is enriched  by a concept $c$, becoming $(S',C_1\cup c)$. This means that the subfield $(S',C_1\cup c)$ is increasing --- as it is a subfield of $(S,C_1)$, it is a subfield increase. In the limit case, when all agents working on $C_1$ are also working on $c$, the superfield $(S,C_1)$ becomes exactly  $(S,C_1\cup c)$. In all other cases, it is $(S',C_1\cup c)$, a strictly smaller subfield of $(S, C_1)$, with $S' \subset S$. 
Conversely, if a field $(S',C_1\cup c)$ is to lose a specific concept $c$, the subcategory $(S',C_1\cup c)$ is going to decrease relatively to $(S,C_1)$.}

\begin{figure}
\begin{center}
\includegraphics[width=7cm%0.92\linewidth%, trim=65 50 0 5
]{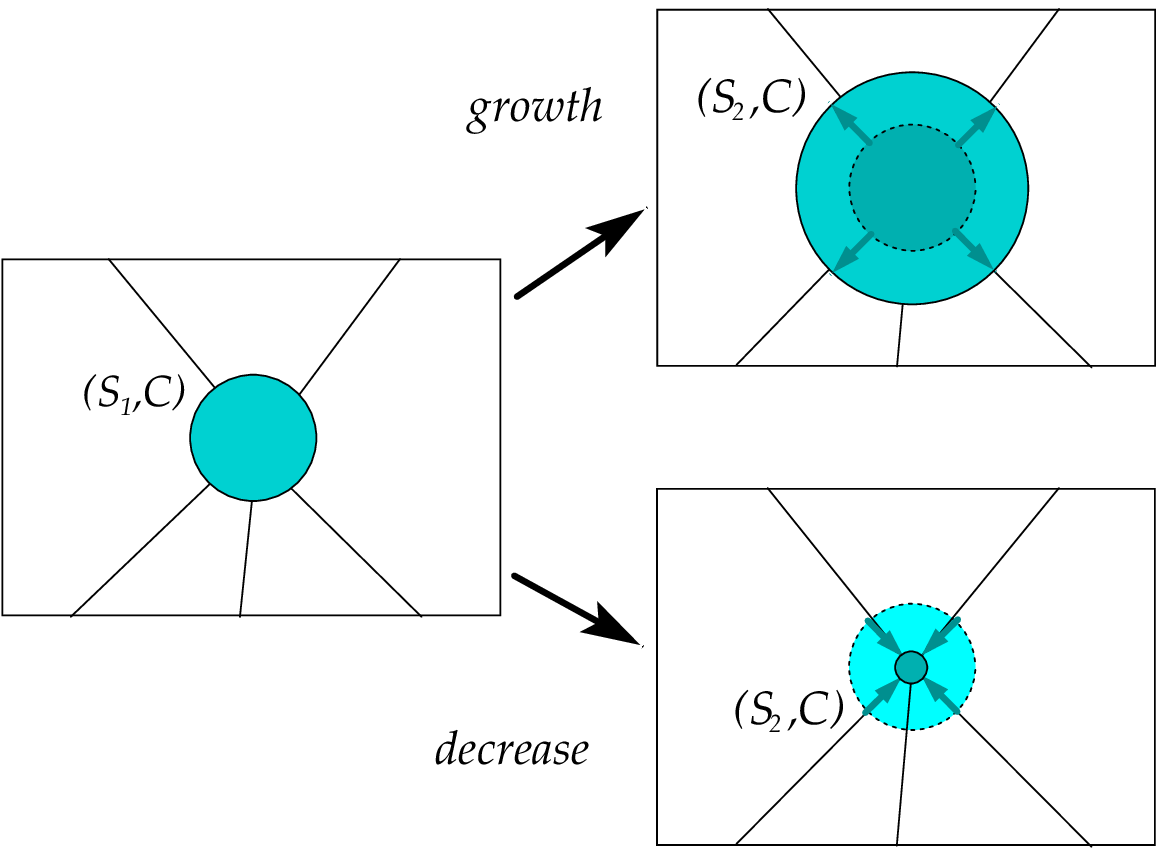}
\vspace{0.45cm}
\hrule
\vspace{0.25cm}
\includegraphics[width=5.62cm%0.74\linewidth%, trim=65 50 0 5
]{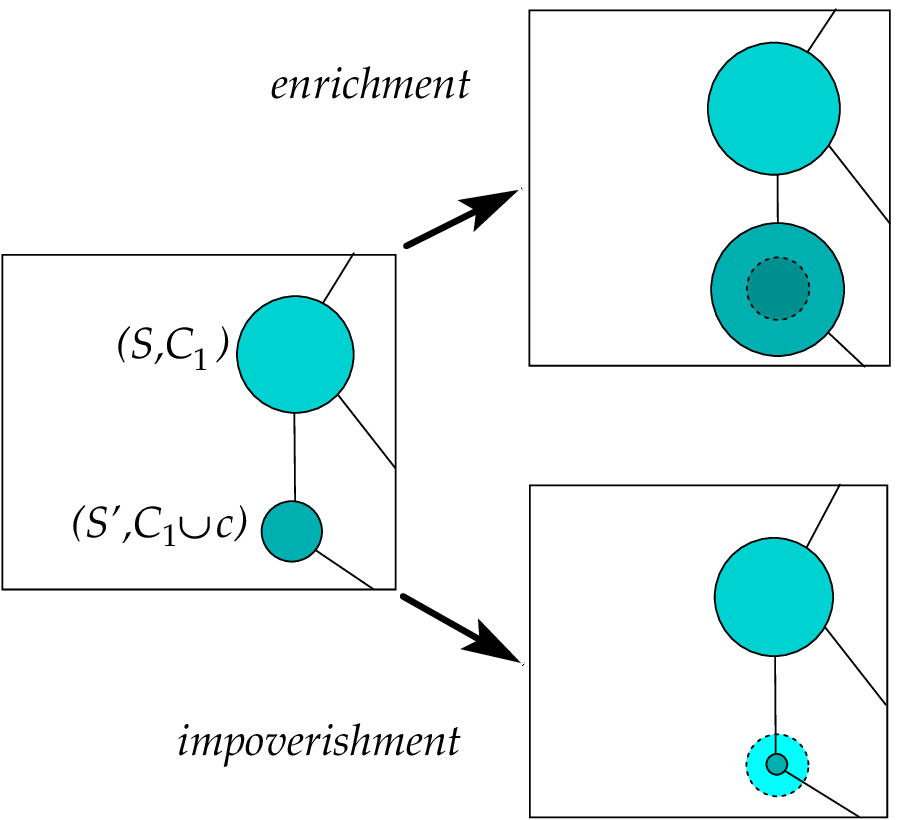}
\vspace{0.25cm}
\hrule
\vspace{0.45cm}
\includegraphics[width=7.14cm%0.94\linewidth%, trim=65 50 0 5
]{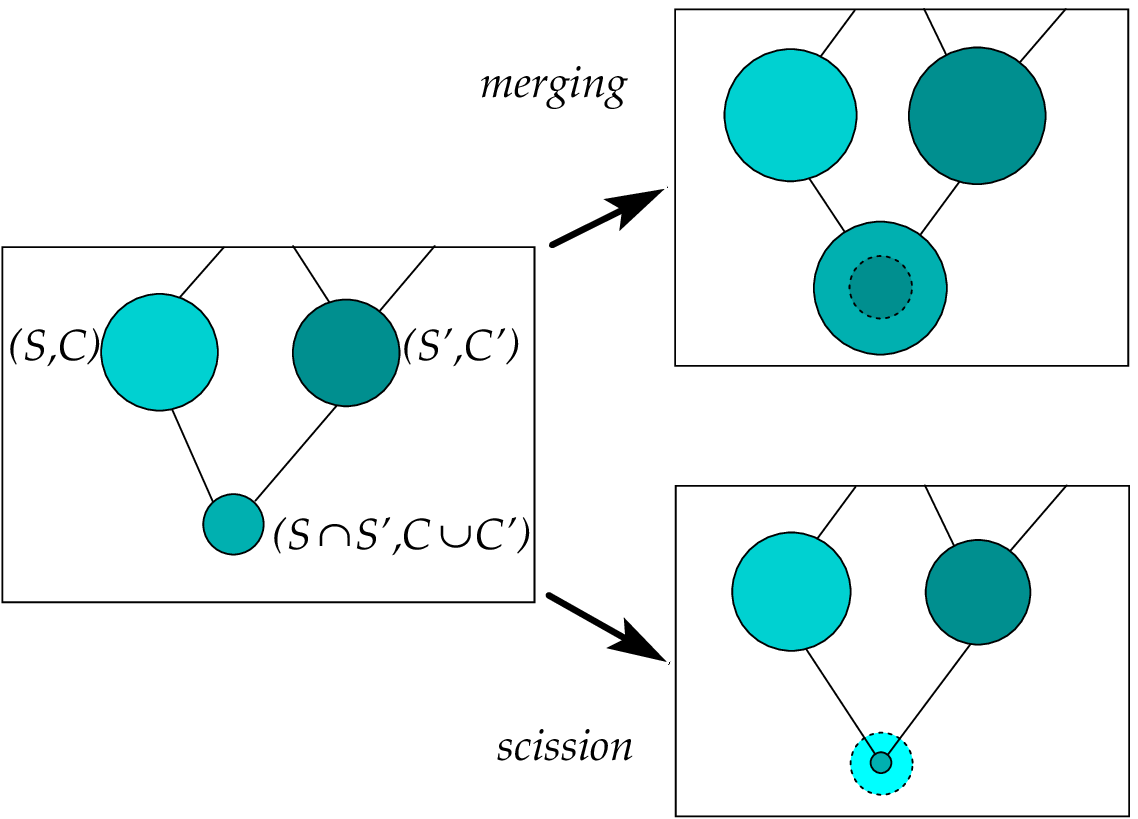}
\end{center}
\caption{\label{fig:dynpatterns} %\small
\emph{Top:} progress or decline of a given EC  $(S_{1}, C)$, whose agent set is growing (above) or decreasing (below) to $S_{2}$. \e{Middle:} enrichment or impoverishment of $(S,C_{1})$ by a concept $c$, through a population change of the subfield \mbox{$(S', C_{1}\cup c)$}. \emph{Bottom:} emergence or disappearance of a joint community (diamond bottom) based on two more general ECs, $(S,C)$ and $(S',C')$. Disk sizes represent agent set sizes.}
\end{figure}

Finally, the union of various fields into an interdisciplinary subfield as well as the scission of this interdisciplinary field comes in fact to an increase or a decrease of a joint subfield ---  geometrically, this means that a diamond bottom is emerging or disappearing (see Fig. \ref{fig:dynpatterns}--bottom).  %(What about  diamond tops? What when the superfield has not the same label as the subfields?).
Obviously a merging (respectively a scission) is also an enrichment (resp. impoverishment) of each of the superfields.

Hence, each of these three kinds of patterns corresponds to a growth or a decrease in agent set size.  The interpretation of the population change ultimately depends on the EC position in the partial lattice, and should vary according to whether (i) there is simply a change in population, (ii) the change occurs for a subfield and (iii) this subfield is in fact a joint subfield.
These patterns, summarized on Fig. \ref{fig:dynpatterns}, describe epistemic evolution with an increasing precision. More precise patterns could naturally be proposed, but as we shall see, these ones are nevertheless sufficiently relevant for the purpose of our case study.

\section{Case study}\label{sec:case}

In this section we detail an empirical protocol for this method and present our findings on a particular case study.

\subsection{Empirical protocol}

%\subsubsection*{Data collection method}

%We aim at building a temporal series of relationships between authors and concepts, from their coappearance within articles. 

To describe the community evolution over several periods of time, we need data telling us \emph{when} an agent $s$ uses a concept $c$.  %(authors linked to concepts) 
To this end,  assuming articles to be a faithful account of what their authors deal with, we use a database of dated articles.

Accordingly, we divide the database into several time-slices, and build a series of relation matrices  aggregating all events of each corresponding period. 
Before doing so, we need to specify the way we choose the \emph{time-slice width} (size of a period), %sample we aggregate data on% to get a static snapshot of a given period
  the \emph{time-step} (increment of time between two periods) % set the scale of the dynamics). 
and the way we \emph{attribute a concept} to an agent, thus  to an article.

%\begin{description}
\paragraph{Time-slice width} We must choose a sufficiently wide time-slice in order to take into account minority communities (who publish less) %hence less frequently) %and all articles written during a given period in order to
 and to get enough information for each author (especially those who publish in multiple fields).%so that these people publish. 
%Small sample sizes exhibit the risk of taking too few articles either for authors that publish more than one article, or for authors or communities that publish less frequently.
\footnote{For instance, extremely few authors publish more than one paper during a 6-month period, so obviously 6-month time-slices are not sufficient. %However, the data shows that out of the first 1000 authors, 75 \% of them have published no more than one paper by the end of 2003.
%also in an exploding field like the one we considered, a lot of new authors appear every year. Remember that we start with a 1000 authors community to arrive at almost 10,000 authors at the end of 2003. 
%In general there are around 1.7 papers per author, should it be a 2-year or a 15-year time-slice, with
%However, in our data 75\% of the authors only publish one paper, should it be a 2-year or a 15-year time-slice. %For these authors obviously a short time-slice is fine, since catching them once is catching their whole intent. However, we do not wish to miss authors that publish more, and ideally on (even slightly) different topics.
}
Doing so also smoothes the data by reducing noise and singularities due to small sample sizes.

However, when taking a longer sample size, we take the risk of merging several periods of evolution into a single time-slice. %So, the time-slice needs not be too large either. 
There is arguably a tradeoff between short but too unsignificant time-slices, and long but too aggregating ones. This parameter must be empirically adapted to the data: depending on the case, it might be relevant to talk in terms of months, years or decades.

\begin{figure}\begin{center}
\includegraphics[width=7.14cm%0.94\linewidth%, trim=65 50 0 5
]{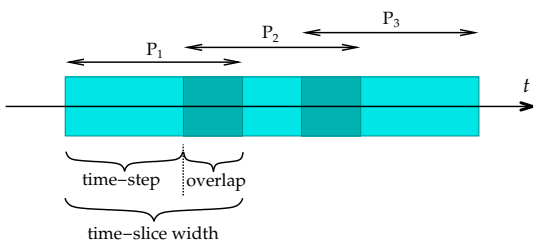}\end{center}\caption{\label{fig:timeslices} Series of overlapping periods $P_{1}$, $P_{2}$ and $P_{3}$.}\end{figure}

\paragraph{Time-step}
The time-step is the increment between two time-slices, so it defines the pace of observation. We need to consider overlapping time-slices, 
since we do not want to miss developments and events  covering the end of a period and the beginning of the next one.  %, If we cut the periods at some point. We do not want this to happen, so
Therefore, we need to choose a time-step strictly shorter than the time-slice width, as shown on Fig. \ref{fig:timeslices}. %In this way, there is an overlap between two successive periods.

Moreover, the time-step is strongly related to the community \emph{%characteristic 
time-scale}:  seeing almost no change between two periods would indicate that we are below this time-scale. We need to pick out a time-step such that successive periods exhibit sensible changes.%POUR LA THESE \footnote{We may introduce a more objective method for choosing time-step and overlap sizes. Consider indeed the density of evolution patterns ``$d(i)=\#\text{patterns during i}/\text{time-slice width}$'',  for a given time-slice $i$. The goal is thus to get the maximum uniformity in time-slice significance, which is equivalent to have the smallest variance for $d$. We could thus draw the variance $\sigma_{d}$ for various values of time-step and overlap, and select the values that yield the smallest variance.}

\paragraph{Concept attribution} 
We attribute to each author the concepts he used in his articles. We thus need to define what kind of concepts we expect to extract from articles. First, considering article keywords might seem to be a relevant and convenient method. However, keywords are poor and heterogenous indicators, since authors often omit important keywords or choose randomly a keyword instead of another. 

We actually consider \emph{each word or nominal group as a concept}, and dismiss more complicated linguistic phenomena such as homonymy, po\-ly\-se\-mia or synonymy.\footnote{More technically, we only consider words chosen from an expert-made selection among the most frequent words, excluding common and rhetorical words (empty words) as well as non-words (figures, percentages, dates, etc.). Then, we do not distinguish morphological variants such as plural, etc. %We also group words from the same very close semantic fields, such as ``brain'', ``nerves'', ``neural'' and ``neuron''.
}
We also proceed with title and abstract only, because complete contents are seldom available. While apparently rough, these minimal assumptions do not prevent us from building significant taxonomies.

\subsection{Case and dataset description}
We considered the particular  community of embryologists working on the model animal \emph{``ze\-bra\-fish''}. This is a clearly defined community, with a decent size. We focused on publicly available bibliographic data from the \T{MedLine} database, covering the years 1990-2003. This timespan corresponds to a recent and important period of expansion for this community, which gathered  approximately $1,000$ agents at the end of 1995, and reached nearly $10,000$ people by end-2003. 
We chose a time-slice width of 6 years, with a time-step of 4 years --- that is, a 2 years overlap between two successive periods.  
We thus splitted the database in three periods: 1990-1995, 1994-1999 and 1998-2003.  %The pictures correspond to the situation at the end of each period.

To limit computation costs, we limited the dictionary to the 70 most used and significant words in the community, selected with the help of our expert. We also considered for each period a random sample of 255 authors. 
 Besides, we used a fixed-size author sample so as to distinguish taxonomic evolutions from the trend of the whole community. Indeed, as the community was growing extremely fast, an EC could become more populated because of the community growth, while it was in fact  becoming less attractive.
With a fixed-sized sample, we could compare the relative importance of each field with respect to others within the evolving taxonomy. 
\subsection{Rebuilding history}

Few changes occured between the first and the second period, and between the second and the third period: the second period is a transitory period between the two extreme periods. This seems to indicate that a 4-year time-step is slightly below the time-scale of the community, %\footnote{Although we do know not of  scientific reports supporting the following fact, according to our expert 
while 8 years can be considered a more significant time-scale.\footnote{Kuhn \shortcite{kuhn:stru} asserts that old ideas die with old scientists --- equivalently new ideas rise with new scientists. In this community, 8 years could represent the time required for a new generation of scientists to appear and define new topics; \mbox{e.g.} the time between an agent graduation and his first students graduation.} 

%\begin{figure}
%\begin{center} 
%\includegraphics[width=1\linewidth, trim=5 50 10 5]{90-95ok.eps}
%\includegraphics[width=1\linewidth, trim=5 50 10 5]{98-03ok.eps}\end{center}

%{\small \emph{Legend: \emph{All: the whole community}, \emph{Hom:} homologue/homologous, \emph{Mou:} mouse, \emph{Hum:} human, \emph{Ver:} vertebrate, \emph{Dev:} development,  \emph{Pat:} pattern,  \emph{Brn:} brain/neural/nervous/neuron,  \emph{Spi:} spinal,  \emph{Crd:} cord,  \emph{Ven:} ventral,  \emph{Dor:} dorsal,  \emph{Gro:} growth,  \emph{Sig:} signal,  \emph{Pwy:} pathway,  \emph{Rec:} receptor.}}
%\caption{\label{fig:lattices} Two partial lattices representing the community at the end of 1995 (\emph{top}) and at the end of 2003 (\emph{bottom}). Figures in parentheses indicate the number of agents per EC. Lattices established from a sample of 255 agents (out of $1,000$ for the first period vs. $9,700$ for the third one).}
%\end{figure}

\begin{figure}
\begin{center} 
\includegraphics[width=1\linewidth, trim=5 50 10 5]{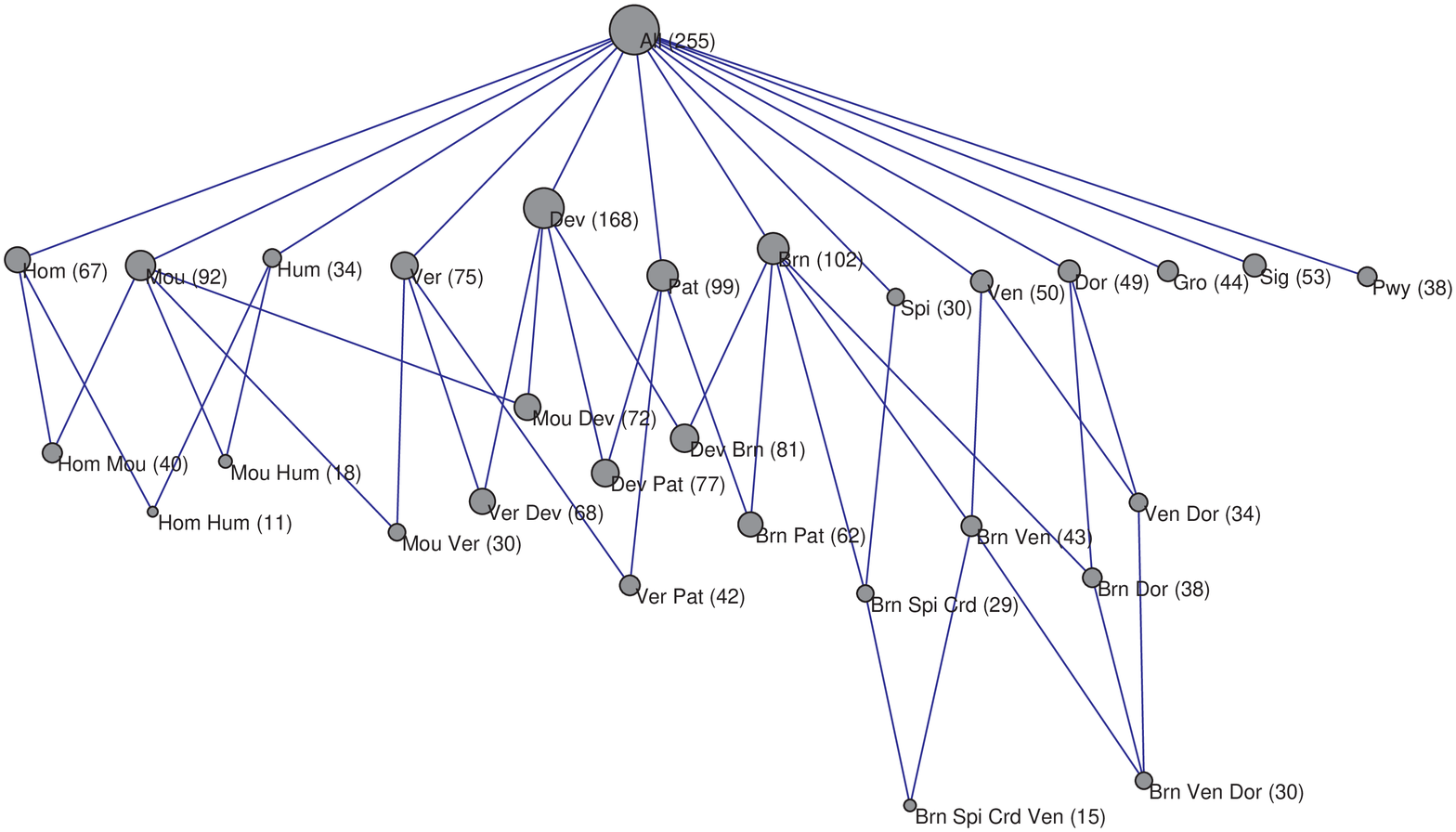}
\includegraphics[width=1\linewidth, trim=5 50 10 5]{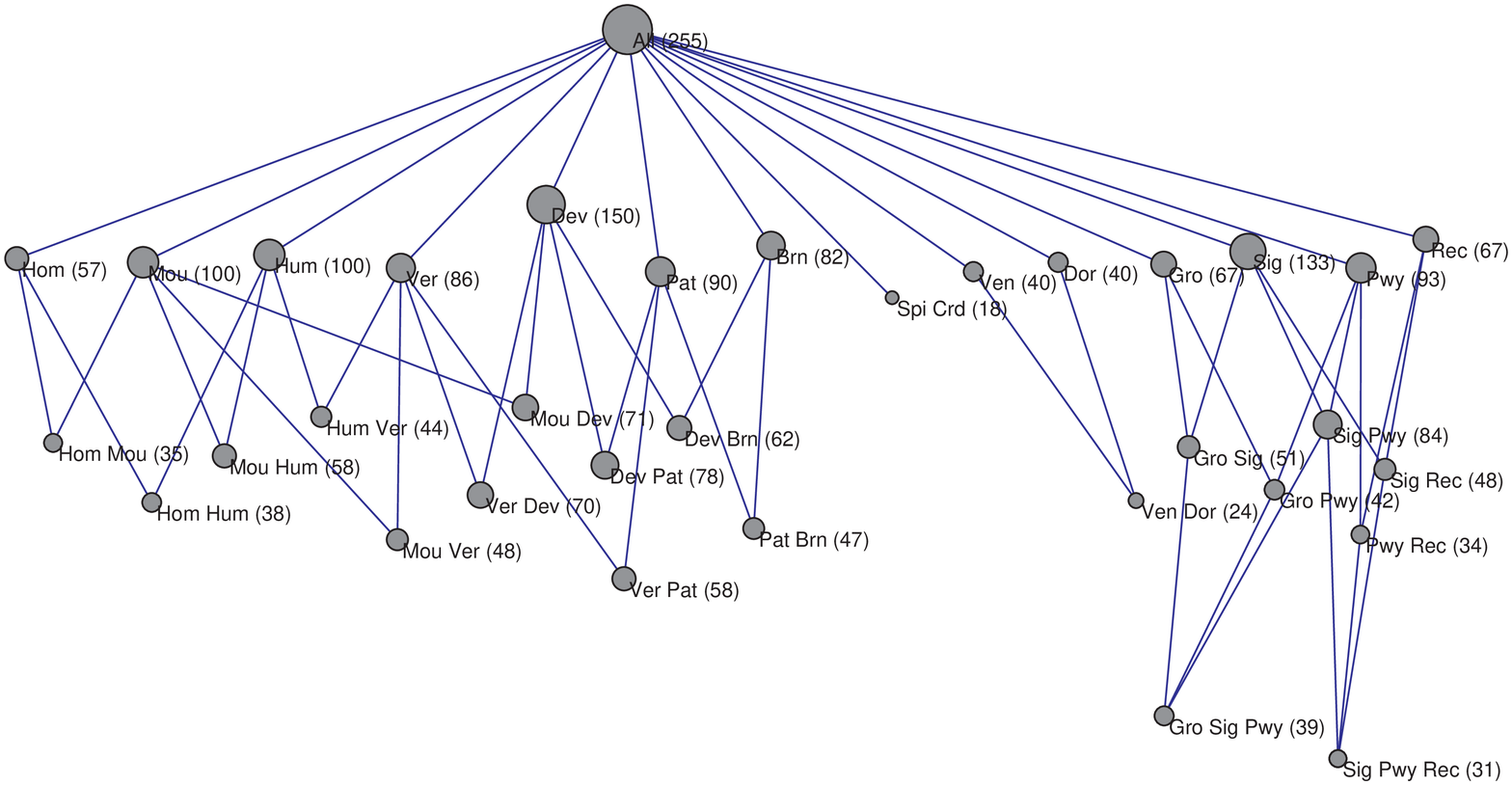}\end{center}

{\small \emph{Legend: \emph{All: the whole community}, \emph{Hom:} homologue/homologous, \emph{Mou:} mouse, \emph{Hum:} human, \emph{Ver:} vertebrate, \emph{Dev:} development,  \emph{Pat:} pattern,  \emph{Brn:} brain/neural/nervous/neuron,  \emph{Spi:} spinal,  \emph{Crd:} cord,  \emph{Ven:} ventral,  \emph{Dor:} dorsal,  \emph{Gro:} growth,  \emph{Sig:} signal,  \emph{Pwy:} pathway,  \emph{Rec:} receptor.}}
\caption{\label{fig:lattices} Two partial lattices representing the community at the end of 1995 (\emph{top}) and at the end of 2003 (\emph{bottom}). Figures in parentheses indicate the number of agents per EC. Lattices established from a sample of 255 agents (out of $1,000$ for the first period vs. $9,700$ for the third one).}
\end{figure}

We hence focus on two periods: the first one, 1990-1995, and the third one, 1998-2003.  The two corresponding partial lattices are drawn on \mbox{Fig. \ref{fig:lattices}} (page \pageref{fig:lattices}). We observe that:
\begin{itemize}
\item {\em First period (1990-1995), first partial lattice:} \{\e{develop}\} and \{\e{pattern}\} strongly structure the field: they are both large communities and present in many subfields. %Alternatively, \e{growth} and \e{homolog} have a certain place. 

Then, slightly to the right of the partial lattice, a large field is structured around \e{brain}\footnote{We actually grouped \e{brain}, \e{nerve}, \e{neural} and \e{neuron} under this term.} and \e{ventral} along with \e{dorsal}. Excepting one agent, the terms  \e{spinal} and \e{cord} form a community with \e{brain}; this dependance suggests that the EC \{\e{spinal}, \e{cord}\} is necessarily linked to the study of \e{brain}. Subfields of \{\e{brain}\} also involve \e{ventral} and \e{dorsal}. In the same view, \{\e{brain, ventral}\} has a common subfield with \{\e{spinal, cord}\}.% while \{\e{brain, dorsal}\} does not: it appears indeed that the spinal cord  stems from the ventral not the dorsal plate.

To the left, another set of ECs is structured around \{\e{homologous}\}, \{\e{mouse}\} and \{\e{vertebrate}\}, and  \{\e{human}\}, but significantly less.% for \e{human}),  both related  to \e{homologous}, but significantly more for \{\e{mouse}\}. 

\item {\em Third period (1998-2003), second partial lattice:} We still observe a strong structuration around \{\e{develop}\} and \{\e{pattern}\}, suggesting that the core topics of the field did not evolve. 

However, we notice the strong emergence of three communities, \{\e{signal}\}, \{\e{pathway}\} and \{\e{growth}\}, and the appearance of a new EC, \{\e{receptor}\}. These communities form many joint subcommunities together, as we can see on the right of this lattice, indicating a convergence of interests.

Also, there is a slight decrease of \{\e{brain}\}. More interestingly, there is no joint community anymore with \{\e{ventral}\} nor \{\e{dorsal}\}.
The interest in \{\e{spinal cord}\} has decreased too, in a larger proportion.

Finally, \{\e{human}\} has grown a lot, not \{\e{mouse}\}. These two communities are both linked to \{\e{homologous}\} on one side, \{\e{vertebrate}\} on the other.
While the importance of \{\e{homologous}\} is roughly the same, the joint community with \{\e{human}\} has increased a lot. The same goes with  \{\e{vertebrate}\}: this EC, which is almost stable in size, has a significantly increased role with \{\e{mouse}\} and especially \{\e{human}\} (a new EC \{\e{vertebrate, human}\} just appeared).

\end{itemize}

To summarize in terms of patterns:
%\begin{itemize}
%\item
some communities  were stable (e.g. \{\e{pattern}\}, \{\e{develop}\}, \{\e{vertebrate, develop}\}, \{\e{homologous, mouse}\}, etc.), some enjoyed a burst of interest (\{\e{growth}\}, \{\e{signal}\}, \{\e{pathway}\}, \{\e{receptor}\}, \{\e{human}\}) or  suffered less interest (\{\e{brain}\} and \{\e{spinal cord}\}). % Qualitatively, \{\e{homologous}\} and \{\e{vertebrate}\} have gained \e{human}, while \{\e{brain}\} lost \{\e{spinal cord}\}. 
Also, some ECs merged (\{\e{signal}\}, \{\e{pathway}\}, \{\e{receptor}\} and \{\e{growth}\} altogether; and \{\e{human}\} both with \{\e{vertebrate}\} and  \{\e{homologous}\}), some splitted (\{\e{ventral-dorsal}\} separated from \{\e{brain}\}). We did not see any \emph{strict} enrichment or impoverishment --- even if, as we noted earlier, merging and splitting can be  interpreted as such.

We can consequently suggest the following story: (i) research on brain and spinal cord depreciated, weakened their link with ventral/dorsal aspects (in particular the relationship between ventral aspects and the spinal cord), (ii) the community started to enquire relationships between signal, pathway, and receptors (all actually related to biochemical messaging), together with growth (suggesting a messaging oriented towards  growth processes), indicating new very interrelated concepts prototypical of an emerging  field, and finally (iii) while mouse-related research is stable, there has been a significant stress on human-related topics, together with a new relationship to the study of homologous genes and vertebrates, underlining the increasing role of \{\e{human}\} in these differential studies and their growing focus on human-zebrafish comparisons (leading to a new ``interdisciplinary'' field).

Point (ii) entails more than the mere emergence of numerous joint subcommunities: all pairs of concepts in the set \{\e{growth, pathway, receptor, signal}\} are involved in a joint subfield. Put differently these concepts form a clique of joint communities, a pattern which may be interpreted as \emph{paradigm emergence} (see Fig.~\ref{fig:lattices}--bottom).

\paragraph{Comparison with real taxonomies} We compared these findings with empirical taxonomical data, coming both from:
\begin{enumerate}
\item Expert feedback: Our expert, Nadine Peyri\'eras, confirms that points (i), (ii) and (iii) in the previous paragraph are an accurate description of the field evolution. For instance, according to her, the human genome sequencing in the early 2000s \cite{inte:init} opened the path to zebrafish genome sequencing%\cite{sangerinstitue}
, which made possible a systematic comparison between zebrafish and humans, and consequently led to the development described in point (iii). In addition, the existence of a subcommunity with \e{brain}, \e{spinal cord} and \e{ventral} but not \e{dorsal} reminded her the initial curiosity around the ventral aspects of the spinal cord study, due to the linking of the ventral spinal cord to the mesoderm (notochord), {i.e.} the rest of the body.

\item Litterature: The only article yet dealing specifically with the history of this field seems to be that of Grunwald \& Eisen \shortcite{grun:head}. This paper presents a detailed chronology of the major breakthroughs and steps of the field, from the early beginnings in the late 1960s to the date of the article (2002). While it is hard to infer the taxonomic evolution until the third period of our analysis, part of their investigation confirms some of our most salient patterns:  \emph{``Late 1990s to early 2000s: Mutations are cloned and several genes that affect common processes are woven into molecular pathways''} --- here, point (ii). Note that some other papers address and underline specific  concerns of the third period, such as the development of comparative studies \cite{brad:smal,dool:zebr}.

\item Conference proceedings: Finally, some insight could be gained from analyzing the evolution of the session breakdown for the major conference of this community, ``Zebrafish Development \& Genetics'' \cite{zebr:proc}. Topic distribution depends on the set of contributions, which reflects the current community interests; yet it may be uneasy for organizers to label sessions with a faithful and comprehensive name --- \e{``organogenesis''} for instance covers many diverse subjects. Reviewing the proceedings roughly suggests that comparative and sequencing-related studies are an emerging novelty starting in 1998, at the beginning of the third period, which agrees with our analysis. On the contrary, the importance of issues related to the brain \& the nervous system, as well as signaling, seem to be constant between the first and the third period, which diverges from our conclusions.

\end{enumerate}

The expert feedback here is obviously the most valuable, as it is the most exhaustive and the most detailed as regards the evolving taxonomy --- the other sources of empirical validation are more subject to interpretation and therefore more questionable. A more comprehensive empirical protocol would consist in including a larger set of experts, which would yield more details as well as a more intersubjective viewpoint, thus objective.  

%Add a word on what is a good taxonomy \cite{voge:qual}.

\section{Discussion}\label{sec:discussion}

%\paragraph{Representing evolving taxonomies}
We are hopeful that this method can be widely used for representing and analyzing static and dynamic taxonomies. 
In the first place, it could be  helpful to historians of science, in domains where historical data is lacking --- notably when examining the recent past. 
Studies such as the recent history of the zebrafish community, written by scientists themselves from the zebrafish community \cite{grun:head}, could profit from such non-subjective analysis. In this particular case  the present article might be considered the second historical study of the ``zebrafish'' community. %, with the reinforcement of the intersubjective viewpoint yielded by the use and analysis of GL evolution. %More generally, agents could be able to know their community extent and structure.
At the same time, with the growing number of publications, some fields produce thousands of articles per year. It is more and more difficult for scientists to identify the extent of their own community: they need  efficient representation methods to understand their community structure and activity. 

\begin{figure}
\begin{center}
\includegraphics[scale=0.9]{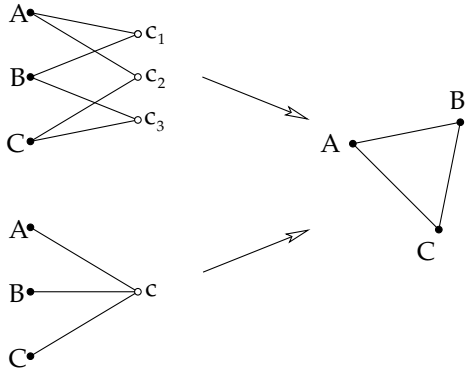}
\end{center}
\caption{\label{fig:projection} Two significantly different two-mode datasets (\emph{left}) yield an identical one-mode projection (\emph{right}), when linking pairs of agents sharing a concept. $A$, $B$, $C$ are agents, $c$, $c_1$, $c_2$, $c_3$ are concepts.}
\end{figure}

More generally, unlike many categorization techniques, community labelling here is straightforward, as agents are automatically bound to a semantic content. Additionally, these categories would have been hard to detect using single-network-based methods, for instance because agents of a same EC are not necessarily socially linked. Moreover,  projection of such two-mode data onto single-mode data often implies massive information loss (see Fig.~\ref{fig:projection}). Finally, the question of overlapping categories --- hardly addressed when dealing with dendrograms --- is easily solved when observing communities through lattices.

Also, using this method is  possible  in at least any practical case involving a relationship between agents and semantic items. As stated by Cohendet, Kirman \& Zimmermann \shortcite{cohe:emer}, {\em ``a representation of the organization as a community of communities, through a system of collective beliefs (...), makes it possible to understand how a global order (organization) emerges from diverging interests (individuals and communities).''\footnote{\emph{``Une repr\'esentation de l'organisation comme une communaut\'e des communaut\'es, \`a travers un syst\`eme de croyances collectives (...), permet (...) de comprendre comment \'emerge un ordre global (organisation) \`a partir d'int\'er\^ets divergents (individus et communaut\'es)''.}}} % POUR LA THESE SEULEMENT: In fine, the partial GL yields a hypergraph of the community, with 
In addition to epistemology, scientometrics and sociology, other fields of application include economics (start-ups dealing with technologies, through contracts), linguistics (words and their context, through co-appearance within a corpus), marketing (companies dealing with ethical values, through customers cross-preferences), and history in general (e.g. evolution of industrial patterns linked to urban centers).%\footnote{We thank Douglas White for providing helpful longitudinal entailment data describing industrial specialization of cities during the middle age% \cite{whit:civi}

%Autre application à réserver à la statique en général: navigation dynamique dans un treillis, à la manière des dynamic taxonomies de Sacco.
 
\paragraph{Lattice manipulation}
%We can think of several drawback as regards reduction, pattern tracking, terminology, dynamics, etc. calling the following improvements:
%\begin{itemize}
%\item

%{\em lattice computation} -- 
On the other hand, our method could  enjoy several improvements. % with respect to lattice manipulation and dynamics study.
Primarily, computing the whole GL then selecting a partial lattice is certainly not the most efficient option. Computing only the lattice part most likely to contain basic-level taxa could  perform better -- using for instance a revised algorithm computing the upper part and its ``valuable'' descendance. 
Similarly, selection heuristics must allow for significant child nodes to appear. Indeed, when two fields do not seem to form a joint subfield in the partial lattice, it is hard to know whether they actually form a joint subfield but are below the threshold. In the second lattice for instance, although of similar importance as \{\e{spinal cord}\} (17 vs. 18 agents), the EC \{\e{brain, spinal cord}\} is excluded by the selection threshold and does not appear, possibly leading us to wrongly deduce that  \{\e{brain}\} does not mix with \{\e{spinal cord}\}.%, while it is only a question of threshold. 

%\item 
%\emph{sublattices within the lattice} -- 
%In order to get better EC representation,
In the same direction, we could endeavor to exclude false positives such as fortuitous intersections (as discussed in section \ref{sec:falsepositives}) and merge clusters of ECs into single multidisciplinary ECs (like for instance ``\e{signal}'', ``\e{pathway}'', ``\e{receptor}''). This would lead to reduced partial lattices containing merged sublattices. Questions arise however regarding the best way to define a cluster of ECs without destroying overlapping communities, one of the most interesting feature of GLs. Accordingly, it could also be profitable to disambiguate and regroup terms in the lattice using for instance Natural Language Processing (NLP) tools \cite{ide:word}: certainly not everyone assigns the same meaning to  \e{``pattern''};  we would thus have to introduce \e{``pattern--1''}, \e{``pattern--2''}, etc. 

Lastly, considering that some authors are more or less strongly related to some concepts, the binary relationship may seem too restrictive. To this end, we could use a weighted relation matrix together with  fuzzy GLs \cite{belo:fuzz}.

%In this respect, are we so sure all nodes in a GL are real communities?

%\item
%{\em threshold of the selection function} -- 

%\item 
\paragraph{Dynamics study}

%\item 
%{\em stability of agent sets} -- %We could also track the evolution of sets of agents. Are agents of a given EC stable? 
Another major class of improvements is related to the study of the dynamics. Indeed, we are now able to represent an evolving taxonomy but we ignore whether individual agents have fixed roles or not.
In particular, the stability of the size of an EC does not imply the stability of its agent set. %There would be no particular problem to extend the present method. %U sing a sample of agents. 
Fortunately, even if our random agent samples are not consistant across periods, it would be easy to rebuild the whole community taxonomy by filling the partial ECs with their corresponding full agent sets. In this case, field scope enrichment or impoverishment could  be described in a better way: by monitoring an identical agent set, and by watching whether its intent increases or not.

%Too hard to observe? it would need us to monitor an agent set (it is dual! what you do is monitor a concept set, and check if there's a change in its extent size) and see whether there is a change in the intent: either it is reduced to small agent sets (uninteresting), or we have a problem since large agent sets are not likely to stay stable, and in this latter case, we need to monitor a ``same'' community and define what a ``same'' community is, across several periods...

More generally, we could address this topic by considering the lattice \emph{dynamics}, instead of adopting a \emph{longitudinal} approach. %,  that is, a comparison of successive static pictures. 
A dynamic study %of GLs instead of a  longitudinal study  %, setting up criteria based on the  lattice dynamics %, instead of just comparing a series of static pictures and patterns therein. 
 would yield a better representation of field  evolution at smaller scales, nevertheless  saving us the empirical discussion about the right time-step.

%\item 
%\paragraph{General questions}
%{\em choice of the terminology} --  
%\item

%\end{itemize}

%voir les montpellierins ? V. Ventos pour le bruit, V. Soldano aussi; ou bien E. Zénou pour les modules + la clôture floue  suggérer dans la thèse l'utilisation de treillis flous

\section*{Conclusion}

%[10 lines max.]
We presented a method for building a   manageable taxonomy, and describing its evolution.  We focused on the structure of epistemic communities, and introduced a formal framework based on Galois lattices to categorize ECs in an automated and hierarchically structured way. Since the resulting lattice is often unwieldy, we proposed  selection criteria for building a partial lattice gathering  the most relevant ECs, in order to get an insightful taxonomy of the community. 
Consequently, the longitudinal study of such partial taxonomies made  possible an historical description. In particular, we proposed to capture stylized facts related to epistemic evolution such as field progress, decline and interaction (merging or splitting). We ultimately applied our method to the subcommunity of embryologists working on the ``zebrafish'' between 1990 and 2003, and successfully compared the results  with taxonomies given by domain experts.

We are  convinced that this method can be  easily improved and fruitfully ported to other domains.

\vspace{.7cm}
%\hrule
\paragraph{Acknowledgements}
\small
The authors wish to express the warmest thanks to Nadine Peyri\'eras for the numerous and essential discussions and advices. We are also grateful of useful comments made by participants of the Sunbelt XXV International Conference on Social Networks. This work has been partly funded by the CNRS.

\vspace{.8cm} \hrule

\bibliography{biblio}
\bibliographystyle{abbrv}
%\end{multicols}
\onecolumn

\end{document}